\documentclass[useAMS,usenatbib]{mn2e}
\usepackage{graphicx,epsfig}

\def\prd{{Phys.~Rev.~D}}  

\title[BAO and non-Gaussianity via QSO clustering ]{Measuring BAO  and  non-Gaussianity via QSO clustering}
\author[U. Sawangwit et al]{U. Sawangwit$^{1}$\thanks{E-mail:
utane.sawangwit@durham.ac.uk (US)}, T. Shanks$^{1}$, S.~M. Croom$^{2}$, M.~J. Drinkwater$^{3}$, S. Fine$^{1}$, 
\newauthor D. Parkinson$^{3}$ \& Nicholas~P. Ross$^{4}$\\
$^{1}$Dept. of Physics, Durham University, South Road, Durham, DH1 3LE, UK\\
$^{2}$Sydney Institute for Astronomy, School of Physics, University of Sydney, NSW 2006, Australia\\
$^{3}$School of Mathematics \& Physics, The University of Queensland, Brisbane QLD 4072, Australia\\
$^{4}$Lawrence Berkeley National Laboratory, One Cyclotron Road, Berkeley, CA 94720, USA}
\begin{document}

\date{Accepted 2011 September 16.  Received 2011 September 16; in original form 2011 June 5}

\pagerange{\pageref{firstpage}--\pageref{lastpage}} \pubyear{2010}

\maketitle

\label{firstpage}

\begin{abstract}
Our goals are (i) to search for BAO and large-scale structure in current
QSO survey data and (ii) to use these and simulation/forecast  results
to assess the science case for a new, $\ga 10\times$ larger, QSO survey.
We first combine the SDSS, 2QZ and 2SLAQ surveys to form a survey of
$\approx60000$ QSOs. We find a hint of a peak in the QSO 2-point correlation
function, $\xi(s)$, at the same scale ($\approx105$h$^{-1}$Mpc) as
detected by \citet{Eisenstein05} in their sample of DR5 LRGs but only
at low statistical significance. We then compare these data with QSO
mock catalogues from the Hubble Volume N-body light-cone simulation used
by \citet{hoyle02} and find that both routes give statistical error
estimates that are consistent at $\approx100$h$^{-1}$Mpc scales. 
Mock catalogues are then used to estimate the nominal survey size 
needed for a 3-4$\sigma$ detection of the BAO peak. We find that 
a redshift survey of $\approx250000$ $z<2.2$ QSOs is required over
$\approx3000$deg$^2$. This is further confirmed by static log-normal 
simulations where the BAO are clearly detectable in the QSO power spectrum 
and correlation function. The nominal survey would on its own produce the 
first detection of, for example,  discontinuous dark energy evolution in
the so far uncharted $1<z<2.2$ redshift range. {\it We further find that a
survey with $\approx50$\% higher QSO sky densities and 50\% bigger area
will give an $\approx6\sigma$ BAO detection, leading to an error
$\approx60$\% of the size of the BOSS error on the dark energy evolution
parameter, $w_a$.}


Another important aim for a QSO survey is to place new limits on
primordial non-Gaussianity at large scales. In particular, it is important
to test tentative evidence we have found for the evolution of the linear
form of the combined SDSS$+$2QZ$+$2SLAQ QSO $\xi(s)$ at $z\approx1.6$, which may
be caused by the existence of non-Gaussian clustering features at high redshift.
Such a QSO survey will also determine the gravitational growth rate at 
$z\approx1.6$ via redshift-space distortions, allow lensing tomography
via QSO magnification bias while also measuring the exact luminosity
dependence of small-scale QSO clustering.

\end{abstract}

\begin{keywords}
quasar clustering
\end{keywords}

\section{Introduction}

Quasi-stellar objects (QSOs) have been used as tracers of large-scale structure for many years
now. The first measurements were made by \citet{osmer81}, then with the
arrival of high-multiplex fibre systems, the subject advanced rapidly
(e.g. \citealt{boyle88,croom05,jose08,ross09} and references therein).
Their clustering at small scales as measured by the correlation function
is known to be consistent with the usual $\gamma=-1.8$ power-law
form for galaxies. The amplitude is comparable to galaxies at low
redshifts and remains reasonably constant with redshift. At larger
scales the power spectrum has been measured to be reasonably consistent
with the standard $\Lambda$CDM cosmological model \citep[e.g.][]{hoyle02,outram03}.

Here, we have combined the largest, spectroscopically confirmed, QSO
surveys from fibre spectrographs including  2QZ \citep{croom04mn}, 2SLAQ
\citep{croom09} and SDSS DR5 \citep{schneider07}   to
form a redshift survey comprising some $\approx60000$ QSOs. We have
already found in these datasets that the small scale clustering of QSOs
is remarkably independent of QSO luminosity at fixed
redshift \citep{shanks11}.

In this paper, we first use the above combined survey to estimate the
large-scale QSO correlation function and search for the BAO feature. We
then outline our initial  motivation for an extended QSO redshift survey
using the `effective survey volume' as a measure of  clustering `grasp'.
We then make an empirical test of the errors on the QSO correlation
function at $\approx100$h$^{-1}$Mpc scales  using both the data and 
mock QSO catalogues from the Hubble Volume N-body simulation. These
routes  allow a first estimate of the survey size needed for a
significant  BAO detection. We  particularly focus on surveys that could
be made with 2dF \citep{lewis02} and AAOmega \citep{smith04} at the AAT
(c.f. the results of  \citet{wang09} for QSO surveys with LAMOST). We
then use static log-normal simulations to test further BAO detectability
and use Fisher matrix and Markov Chain Monte Carlo (MCMC) methods to
test a QSO survey's competitiveness against other routes to the BAO
scale and the evolution of $w$. Finally, we look at the prospects for
detecting evidence for non-Gaussian clustering at large scales via a QSO
survey, in particular focussing on the possibility that the current QSO
surveys show evidence for evolution in the linear regime of clustering,
which might represent evidence for non-Gaussianity, if confirmed in a
larger survey.

\section{QSO Clustering data}

We start by making a study of the results from current QSO surveys
(a) to see if the BAO peak can be detected in the correlation function
and (b) to measure  the errors  to base an empirical estimate of the 
new survey size needed for an accurate BAO measurement. This survey size estimate
will then be compared to those from simulations to determine the best
QSO survey strategy in terms of area and magnitude limit.

\subsection{SDSS, 2QZ and 2SLAQ  surveys}
Previously \citet{croom05} used the 2QZ survey to estimate the QSO
correlation function and its dependence on redshift and luminosity. This
survey contained $\approx22655$ QSOs in two $\approx375$deg$^2$ NGC $+$
SGC strips. The magnitude limit was $18.25<b_J<20.85$ and the resulting
QSO sky density was $31$deg$^{-2}$. \cite{croom05} measured
$s_0=5.4^{+0.42}_{-0.48}h^{-1}$Mpc and $\gamma=1.2\pm0.1$ at
$1<s<25$h$^{-1}$Mpc for the amplitude and slope of the $z$-space 
correlation function, $\xi(s)$.

\citet{jose08} then used the 2SLAQ survey of 9418 QSOs based on SDSS
imaging to test the luminosity dependence of the QSO clustering. The
magnitude limit was $20.5<g_{AB}(\approx b_J)<21.85$ and 
the resulting QSO sky density was $\approx45$deg$^{-2}$, including
the 2dF QSOs where the two surveys overlapped, in a total area of
$\approx200$deg$^2$.  \citet{jose08} found a $\xi(s)$ amplitude and
slope similar to that for 2QZ.

Most recently, \citet{ross09} have analysed the clustering of 30239 QSOs
in the 4013deg$^2$ SDSS DR5 survey to  $i_{AB}=19.1$. This
gives a QSO sky density of 7.5deg$^{-2}$ in the uniform sample where
\citet{ross09} measured $s_0=5.95\pm0.45h^{-1}$Mpc and
$\gamma=1.16_{-0.16}^{+0.11}$ in the $1<s<25$h$^{-1}$Mpc range. 

\subsection{Large-scale clustering comparison}

Here, the clustering analysis has been re-done to use consistent bins at
comoving separations corresponding to  BAO scales in the 2QZ, 2SLAQ and
SDSS-DR5 spectroscopic QSO samples. The data and random catalogues are
the same as those used in the analyses of \citet{jose08} and
\citet{ross09} for 2QZ$+$2SLAQ and SDSS-DR5, respectively. We used the
`UNIFORM' sample with $0.3\le z \le 2.2$ of \citet{ross09} which
contains 30239 QSOs over 4013 deg$^2$. The 2QZ$+$2SLAQ sample contains
28790 $0.3<z<2.9$ QSOs and its small- and intermediate-scale clustering
analyses have been performed by \citet{jose08} (see also \citet{croom05}
for  the 2QZ-only clustering analysis). We perform a new clustering
analysis by counting pairs at separation, $s$, independently for the SDSS
and 2QZ$+$2SLAQ samples. The data-random, $DR$, and random-random, $RR$,
pairs for each sample are normalised by $N_{rd}$ and $N_{rd}^2$,
respectively, where $N_{rd}$ is the ratio between numbers of randoms and
data ($\approx 20$ for the 2QZ$+$2SLAQ and $\approx 30$ for the SDSS
samples). The \citet{ls93} estimator is then used to determine the
$\xi(s)$ from the summed pairs over the different samples. Note that the
results are in good agreement with those using the \citet{hamilton93}
estimator.

\begin{figure}
\hspace{-3mm}\includegraphics[scale=0.5]{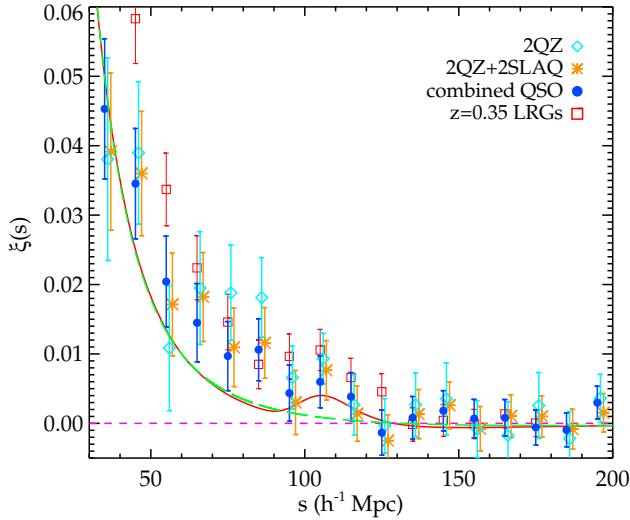} 
\caption{The large-scale redshift-space correlation function results
from the 2QZ, the 2QZ+2SLAQ and the combined SDSS DR5+2QZ+2SLAQ QSO
surveys. The errors are empirically scaled by the average (1.2$\times$)
ratio of jack-knife to Poisson errors in this separation range. The
results are compared to  $\Lambda$CDM (red solid) and no-wiggle (green
dashed) models and also the $z=0.35$ LRG result of \citet{Eisenstein05}.
}
\label{fig:xis_lin}
\end{figure}

In Fig. \ref{fig:xis_lin} we then compare the large-scale clustering
results from the three surveys directly with each other using the
redshift-space correlation function, $\xi(s)$. The cosmology assumed  in
all cases is $\Omega_\Lambda=0.7$, $\Omega_m=0.3$. Fitting the $\xi(s)$
results consistently in the $1<s<30$h$^{-1}$Mpc range, \citet{shanks11}
fitted real-space correlation function scale-lengths, $r_0$, assuming
power-law slope, $\gamma=1.8$, infall parameter, $\beta=0.4$, and
line-of-sight pairwise velocity dispersion $<w^2>^{1/2}=750$kms$^{-1}$.
These authors  found agreement at the $1.4\sigma$ significance level
in these results with SDSS giving $r_0=6.30\pm0.3$h$^{-1}$Mpc, 2QZ
giving $r_0=5.75\pm0.25$h$^{-1}$Mpc and 2SLAQ giving
$r_0=5.70\pm0.35$h$^{-1}$Mpc. The best overall fit is
$r_0=5.90\pm0.14$h$^{-1}$Mpc. Thus  the small-scale results suggest that
it is reasonable to combine these 3 surveys and the $\xi(s)$ result at
large-scales is also shown in Fig. \ref{fig:xis_lin}. The data is compared
to a LCDM model and a `wiggle-free' version of the model \citep{eis_hu98}. 
These models are normalised to match the data at $s=10-30 $h$^{-1}$\,Mpc 
(see Fig. \ref{fig:logxi} below). The errors are
based on Poisson errors calibrated by jack-knife errors in all 3 cases.
Generally the SDSS has the biggest errors, particularly at the smaller
scales. This is because of its relatively low sky density. This means
that at $\approx100$h$^{-1}$Mpc the 2QZ (and 2SLAQ) surveys dominate the
statistics at the predicted BAO scale. The 2QZ survey error is
$\approx2\times$ smaller than the SDSS error and $1.8\times$ smaller
than the 2SLAQ error. The blue filled circles then represent the overall
2QZ$+$2SLAQ$+$SDSS result, produced  by simply adding the QSO-QSO and
QSO-random pairs across the surveys. The resulting error is
$\approx2\times$ larger than the error from the SDSS LRG sample of
\citet{Eisenstein05}. We see that there is some hint of a detection at
105h$^{-1}$Mpc but there is a similarly sized feature at
$\approx85$h$^{-1}$Mpc. Clearly while the data appear promising in terms
of detecting the BAO feature, a larger sample size is needed and in
Section \ref{sect:surveysize} below  we will  translate these empirical errors into a
required survey size to measure the BAO scale. We shall also compare
with the errors predicted by simulations and use these to optimise  the
properties of a new QSO redshift survey for large-scale clustering.

\section{QSO survey effective volume }
\label{sect:veff}
We next estimate the efficiency of future QSO clustering surveys  via their
effective volume ($V_{eff}$). Although  QSO sky densities are lower than for
galaxy surveys, their clustering amplitude is relatively high, the
exposure time required to establish redshifts is generally quite short,
contamination rates are increasingly low and the volumes probed are very
large. Moreover, the low QSO sky density can be viewed as an advantage in
that it may well match the fibre density currently available from
instruments like AAT 2dF.

\begin{figure}
\includegraphics[scale=0.5]{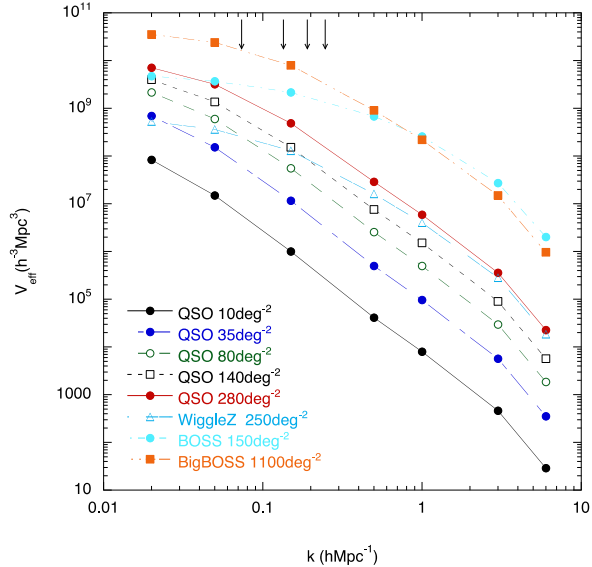} 
\caption{The effective volume as a function of wavenumber of spatial
scale for a $z\approx1.5$ QSO survey of area 3000deg$^2$ and QSO sky
densities varying from 10deg$^{-2}$ to 280deg$^{-2}$ as shown. The
assumed QSO correlation function amplitude was $s_0=6h^{-1}$Mpc.  Also
shown is the effective volume for the BOSS LRG survey at $z\approx0.5$
assuming $s_0=10$h$^{-1}$Mpc and 10000deg$^2$ area. The WiggleZ ELG
survey at $z\approx0.5$ has $s_0=4.4$h$^{-1}$Mpc and 1000deg$^2$ area.
The proposed BigBOSS ELG survey at $z\approx1$ has $s_0=4.4$h$^{-1}$Mpc
and 14000deg$^2$ area. The arrows indicate the positions of the 1st, 
2nd, 3rd and 4th BAO peaks (from left to right).}
\label{fig:veff}
\vspace{-5mm}
\end{figure}

A rough measure of  a survey's capability for  power spectrum or 2-point
correlation function analysis, the effective volume is defined by
\citet{Eisenstein05} and represents the survey volume that has a high
enough QSO density for  the shot noise to lie below the amplitude of a
spatial power spectrum feature such as a BAO oscillation scale. The
power spectrum or correlation function error at a particular scale then
is proportional to $V_{eff}^{-1/2}$. In Fig. \ref{fig:veff} we have
calculated the effective volume for QSO surveys assuming the SDSS DR5
QSO $n(z)$ in the redshift range $0.5<z<2.2$ (see also Fig. 2 of
\citealt{wang09}). We have chosen a nominal survey area of 3000deg$^2$;
effective volumes of other survey  areas will scale linearly. Since the
QSO $n(z)$ is approximately independent of survey magnitude limit, the
main other survey parameter is QSO sky density. We have calculated the
effective volume at  sky densities approximating those for the SDSS, 2QZ
and 2SLAQ surveys at 10, 35 and 80deg$^{-2}$. We also present the
effective volumes at 140deg$^{-2}$, which is approximately the largest
density accommodated by the 2dF fibre positioner (if tiling overlaps 
is considered, see later), and 280deg$^{-2}$
which is the highest QSO density that is available from the Hubble
Volume simulations (see Sect. \ref{sect:hvsim}). The assumed QSO
correlation function amplitude was $s_0=6h^{-1}$Mpc which is also found
to be almost independent of survey limit (see eg \citealt{shanks11}).
From Fig. \ref{fig:veff} we see that QSO  effective volume generally
drops sharply as the spatial wavenumber increases; this drop is at a
faster rate than for more highly sampled galaxy surveys such as WiggleZ.

However, even at the 2SLAQ sky density of 80deg$^{-2}$, we see that at
the scale of the first acoustic peak at $k\approx0.02$hMpc$^{-1}$, the
effective volume of our nominal 3000deg$^2$ survey overtakes that of the
current leading galaxy BAO survey, WiggleZ \citep{blake11}, by a factor
of $\approx3$. Of course, even if the effective volume is only merely
competitive with WiggleZ volume  as it is at the 2nd and 3rd peak
positions, then this still represents an advance, given the
$\approx3\times$ higher redshifts of the QSOs than the WiggleZ galaxies.
To reach the same effective volume at the first acoustic peak of the
current BOSS LRG survey at $z\approx0.55$, a higher QSO sky density of
140deg$^{-2}$ would be needed. This sky density would be reached at
$g\approx22.7$\footnote{We assume that the QSO clustering and bias
continue to be luminosity independent at this limit \citep{shanks11}.},
assuming a $10^{0.3m}$ QSO number count slope \citep{boyle88}. This  QSO
effective volume   again  applies at an $\approx3\times$ higher redshift
than the BOSS LRGs. Even the BigBOSS ELG survey will  produce an
effective volume which is only $\approx2\times$ bigger than for  a
140deg$^{-2}$ QSO survey when renormalised to the same area of sky.
BigBOSS also has a significantly lower average redshift, $z\approx1$.
Thus the relatively crude effective volume measure suggests that a QSO
survey of nominal area 3000deg$^2$ and sky density in the range
80-140deg$^{-2}$ should produce large-scale clustering results that have
similar precision at the first acoustic peak to the BOSS LRG survey but
at significantly higher redshift.

\section{BAO Search in simulations}
\subsection{Hubble Volume}
\label{sect:hvsim}
We next use the Hubble Volume simulation to measure the correlation
function errors directly in mock surveys that range up to higher QSO sky
densities than 2QZ. This simulation \citep{evrard02} used an initial
mass power spectrum with $\Omega_{\rm b}=0.04$, $\Omega_{\rm CDM}=0.26$,
$\Omega_\Lambda=0.7$, $H_0=70$ km s$^{-1}$ Mpc$^{-1}$ and
$\sigma_8=0.9$. Mock QSO catalogues from this simulation were generated
by \citet{hoyle02}. We did consider using newer simulations but although
these frequently had higher resolution, they generally did not have
sufficient volume to accommodate even the original 2QZ survey. The
Hubble Volume mocks are made in the form of past light cones as needed
for  accurate modelling of the QSO survey. The mock QSOs bias relative
to the mass was modelled according to the algorithm of \citet{hatton98}.
The main change from the previous 2QZ mocks is that the QSO sky density
is approximately doubled from 35deg$^{-2}$ to 75deg$^{-2}$. The area of
the mock survey is $15\times75$deg$^2$ or 1.5$\times$ the area of the
2QZ survey. Previously we have shown that our correlation function and
power-spectrum estimation techniques can accurately retrieve the  input
functions in real  and  redshift space. The errors are jack-knife
estimates based on 60 sub-samples from this contiguous area. The
amplitude of the correlation function at small scales is
$r_0=6$h$^{-1}$Mpc, close to the $r_0=5.9$h$^{-1}$Mpc shown by the data.

\begin{figure}
\hspace{-3mm}\includegraphics[scale=0.5]{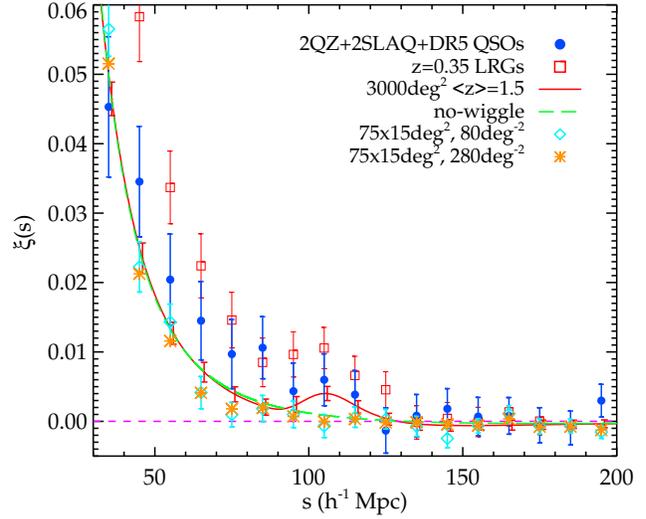}
\caption{Correlation function from QSO Hubble Volume mocks at 35, 80 
and 280deg$^{-2}$ compared to $\Lambda$CDM  and no-wiggle models and
also the observed data. The errors found for the 80deg$^{-2}$ case have
been scaled to a 3000deg$^2$ survey and applied to the $\Lambda$CDM model (red
points $+$ line).}
\label{fig:xis_lin_hv}
\vspace{-5mm}
\end{figure}

Fig. \ref{fig:xis_lin_hv}  shows  the large-scale correlation functions from
the mocks with 1125deg$^2$ area and 80 and 280deg$^{-2}$ QSO sky
densities. We find that the errors are compatible with those
extrapolated using simple Poisson scaling of the data-data pairs from
the 2QZ$+$2SLAQ survey with its smaller QSO sky density and area. In
fact, we find that in the relatively small area of the Hubble Volume
simulation the BAO peak is barely detected, in either the mock at the
standard QSO sky density of 80 deg$^{-2}$ or even at  280 deg$^{-2}$. In
the $\xi(s)$ measured for the unbiased mass (not shown), the feature is 
detected but only at low significance, 1-2$\sigma$. Thus although no
feature is detectable in the relatively small Hubble Volume mocks, these
data can still be used to estimate the likely errors in the
$\approx3\times$ bigger 3000deg$^2$ nominal QSO survey considered in Section
\ref{sect:veff} above.

Fig. \ref{fig:error} shows the ratio of the jack-knife errrors (60
sub-samples) from the above QSO mock catalogues from the Hubble Volume
simulations. The ratio of the errors agrees with the Poisson prediction
between the 35deg$^{-2}$ and 80deg$^{-2}$ (also 105deg$^{-2}$ see later) 
sky densities but the increase to 140deg$^{-2}$ and 280deg$^{-2}$ only 
achieves a factor of 3.3 and 5 improvement in the error rather than the 
Poisson predicted factor of 4 and 8, respectively. So Poisson scaling 
works as far as the sky density of $\approx100$deg$^{-2}$.

\vspace{-3mm}
\subsection{Log-normal simulations}
\label{sect:lnsim}

We also ran static simulations similar to the Gaussian simulations of
\citet{blakekgb03, kgb05}, drawing 3-D mode amplitudes according to
power spectra for a standard $\Lambda$CDM model. The distribution used
for these realisations was up-dated to log-normal rather than Gaussian
to mimic better the effects of non-linearities in the matter
distribution \citep{coles91, blake11}. Fig. \ref{fig:lognormal}a shows
the $P(k)$ analysis of 400  simulations for a 3000 deg$^2$, $z<2.2$ QSO
survey with sky density 90deg$^{-2}$ and a uniform space density of
$n=1.6\times10^{-5}$ h$^3$Mpc$^{-3}$.  We split the redshift range into
3 parts, $0.4<z<1.0$, $1.0<z<1.6$ and $1.6<z<2.2$ with QSO sky densities
of 18, 33 and 39 deg$^{-2}$ with an infall parameter of $\beta = 0.58,
0.43, 0.32$ and $b=1.4, 2.1, 3.0$ in successive redshift ranges
\citep{croom05}. Although the sample is dominated by shot noise, BAOs
are detectable in $P(k)$ in the 2nd and 3rd slices with a precision
comparable to WiggleZ and SDSS-LRGs. In Fig. \ref{fig:lognormal}a, the
dotted lines are the result of an effective volume calculation for the
errors, which agrees well with the scatter in the lognormal
realizations.  The accuracy in the $P(k)$ BAO comes from fitting the
simple \citet{blakekgb03, blake06} model to the realizations. We detect the BAO
in the $1<z<1.6$ bin with $\pm5$\% accuracy for the BAO scale and in the
$1.6<z<2.2$ bin with $\pm3.7$\% accuracy. Overall the BAO scale accuracy in
the $1<z<2.2$ range and $\approx70$deg$^{-2}$ sky density is $\pm3$\%. 

Fig. \ref{fig:lognormal}b shows the mean correlation function result
integrated over the full $0.4<z<2.2$ redshift range. Again we see the BAO
feature clearly detected at $\approx3\sigma$ relative to zero signal.

\section{QSO Dark Energy Survey} 
\label{sect:surveysize}
\subsection{Empirical survey parameter estimates}
\label{sect:empred}
We first take the empirical, jack-knife  error estimates at
100h$^{-1}$Mpc from the 2QZ $\xi(s)$ result, also shown in Fig.
\ref{fig:xis_lin}. Ignoring the 2SLAQ survey simplifies the scaling of
errors since  in the Northern Cap at least, 2SLAQ relied on 2QZ to
supply the QSOs with $b_J<20.85$. We can then simply scale the errors
from 2QZ assuming a sky density of 35deg$^{-2}$ and an area of
750deg$^2$. The 2QZ amplitude and error in Fig. \ref{fig:xis_lin}
suggest that a BAO peak at $\approx100h^{-1}$Mpc will appear  at the
$\approx1\sigma$ level in a QSO sample of the current size ($0.86\sigma$
against the no-wiggle-model and $1.1\sigma$ against zero correlation signal). Hence a
$\approx4\sigma$ detection  will require either a $16\times$ bigger
survey at the 35deg$^{-2}$ density or a $4\times$ bigger survey at
double the QSO sky density.  The simulation results in Fig.
\ref{fig:error} suggest that this Poisson sky density scaling continues
at least as far as $\approx 100$deg$^{-2}$ but not as far as 280deg$^{-2}$. Fig.
\ref{fig:veff} also suggests that the error is $3\times$ rather
than $4\times$ smaller at a QSO sky density of 140deg$^{-2}$ compared to
35deg$^{-2}$, so a survey of only $\approx$1300deg$^{2}$ would be
required to achieve a $4\sigma$ BAO detection. By mainly using previous
2QZ and 2SLAQ survey areas and taking tiling overlaps of $\approx20$\%
into account (to ensure survey areal completeness), the sky density for
new QSO targets might only be $\approx100$deg$^{-2}$, or $\approx300$
per 2dF field. If these QSOs could be efficiently detected with a
contamination of only 25\% or less then it may be possible to achieve
this density with only a single 2dF pointing per field. 2SLAQ achieved a
44\% star contamination rate based on SDSS single epoch imaging data and
improving on this rate mostly depends on achieving improved $ugriz$
photometry compared to SDSS. This should be possible using new surveys
such as VST ATLAS (Shanks et al. 2011, in prep.). Note also that 2SLAQ
used more traditional colour cuts, and methods such as KDE \citep[e.g.][]{richards09}, 
extreme-deconvolution etc., would improve target selection considerably.

\subsection{Simulated survey parameter estimates}
\label{sect:hvlnpred}

\begin{figure}
\hspace{-3mm}\includegraphics[scale=0.47]{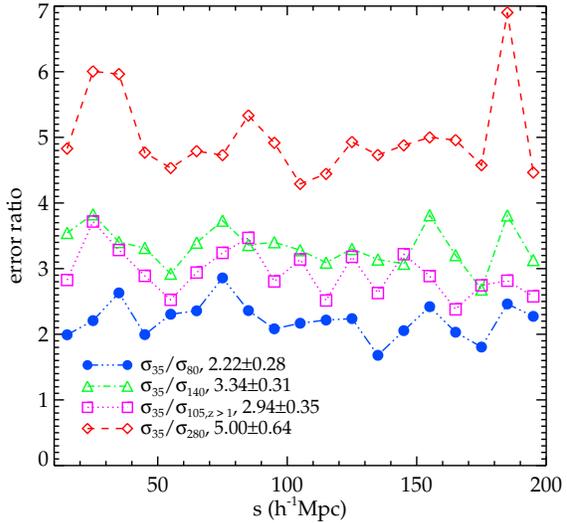}
\caption{The ratio  of the jack-knife errors (60 sub-samples) from the QSO
mock catalogues from the Hubble Volume N-Body simulation, using
densities of 35deg$^{-2}$, 80deg$^{-2}$ 105deg$^{-2}$, 140deg$^{-2}$ and 
280deg$^{-2}$. The final ratios of $2.22\pm 0.28$, $2.94\pm 0.35$, $3.34\pm 0.31$ 
and $5.0\pm 0.64$ can be compared to the QSO sky density ratios of 2.29, 3.0, 4.0 
and 8.0, confirming that the error scales as expected from Poisson statistics 
between 35 and $\approx100$deg$^{-2}$.}
\label{fig:error}
\end{figure}

\begin{figure}
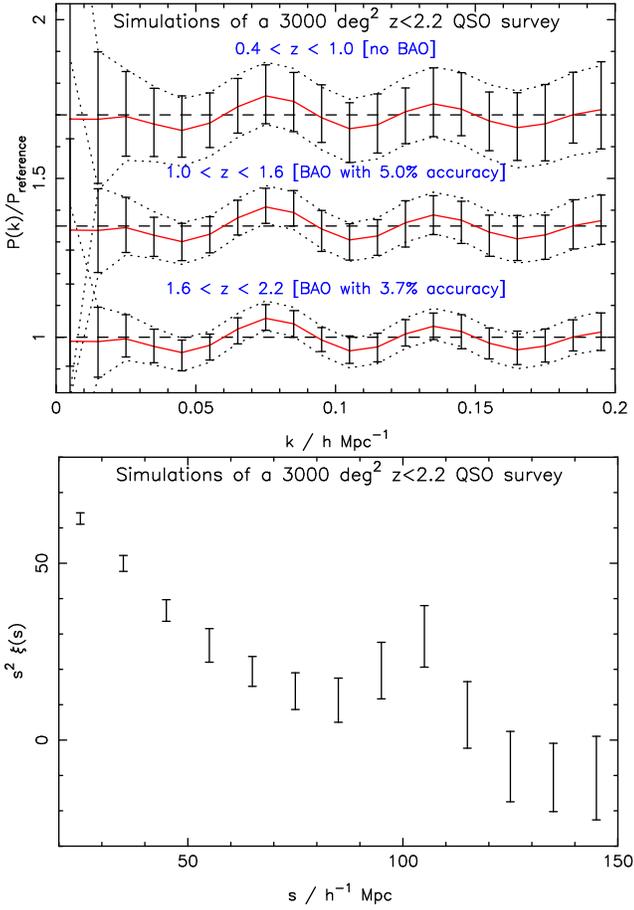

\epsfig{file=pk_qso_new.ps,width=6.0cm,angle=-90}
\epsfig{file=xicomb_qso.ps,width=6.0cm,angle=-90}
\caption{\small (a) Predicted survey sensitivity for BAO from 400 log-normal
simulations. Each successive set of data has been offset in y for
clarity. QSO Power-spectrum BAO accuracy is 3\% over full $1<z<2.2$ range
in nominal survey. (b) The QSO $\xi(s)$ from the log-normal simulations
integrated over the $0.4<z<2.2$ redshift range. The BAO signal is clearly
detected.
}
\label{fig:lognormal}
\end{figure}

We then take the jack-knife based error estimates from the Hubble Volume
simulation. As noted above, we have checked that the jack-knife errors
reduce approximately linearly as the QSO density increases up to
$\approx80$deg$^{-2}$. Fig. \ref{fig:error} shows that as the mock QSO
sky density rose from 35 to 80, 140 and then 280 deg$^{-2}$, factors of 2.3, 
4 and 8.0, the jack-knife error in 60 subsamples reduced by factors of
$2.22\pm0.28$, $3.34 \pm 0.31$ and $5.0\pm0.64$. If we drop the $z<1$ QSOs 
(see below) in the 140deg$^{-1}$ mock catalogue, the sky density becomes 
105deg$^{-1}$ and the error on the $\xi(s)$ only increases by 10\%. Care was taken 
here that the small scale correlation function amplitude remained constant at the 
three sky densities so that the effect of sky density could be easily measured.
These ratios also reasonably agree with taking ratios of the average of
the square root of the three lowest $k$ effective volumes in Fig.
\ref{fig:veff}. So we again conclude that at least up to a sky density of 
$\approx 100$deg$^{-2}$, doubling the sampling rate approximately halves the 
error. We also see that for the 80deg$^{-2}$ QSO mock, extrapolating the 
error from the mock survey area of 1125deg$^2$ to 3000deg$^2$, indicates 
that the error is reduced to $\approx27$\% of the error in the 2QZ $\xi(s)$
results of Fig. \ref{fig:xis_lin}, again indicating that these survey
parameters will produce an $\approx4\sigma$ detection of the BAO feature
in $\xi(s)$. Finally, the error on the $\xi(s)$ peak at
$\approx105$h$^{-1}$Mpc in the log-normal simulations in Fig.
\ref{fig:lognormal}b represents again an $\approx3-4\sigma$ detection. So
there is generally excellent agreement between simulated
estimates and the empirical results from Section \ref{sect:empred}
that a 3000deg$^2$ QSO survey with an 80-90deg$^{-2}$ QSO sky
density will produce a significant detection of the BAO peak with a
scale measurable to $\approx\pm3$\%.

\begin{figure}
\epsfig{file=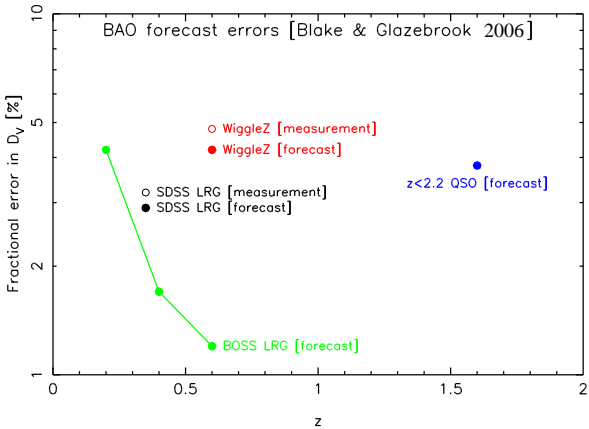,width=8.0cm}
\caption{\small Predicted sensitivity  for BAO in the
nominal QSO survey consistently compared  with other surveys.  QSO
power spectrum BAO accuracy forecasts  from the fitting formula of \citet{blake06}.
}
\label{fig:dv}
\end{figure}

\subsection{Comparison with other surveys}

Fig. \ref{fig:dv} compares the combined $1<z<2.2$ BAO error  for the
dilation scale, $D_V(z)=(D_A(z)^2 cz/H(z))^{1/3}$, of a nominal
3000deg$^2$, 90deg$^{-2}$ QSO survey with other current BAO surveys. The
dilation scale is a measure which combines the information in  the
comoving angular diameter distance, $D_A(z)$, and the Hubble parameter,
$H(z)$. The errors are now generated using the fitting formula of
\citet{blake06} which is calibrated by lognormal realizations
\citep{kgb05} for the BAO measurements. We see that the $\approx\pm4$\%
error from the QSO survey is comparable to the BAO  error  at $z=0.35$
from the SDSS LRG survey and WiggleZ at $z=0.6$. Note that the error
derived  for the dilation scale from the QSO survey-specific
lognormal simulations described above is $\pm3$\% which is clearly the
more directly measured result.  The  results in Fig. \ref{fig:dv}, on
the other hand, have the advantage that they are measured consistently
between the various surveys.

In response to a request from a referee, we note that the Hobby-Eberly
Telescope Dark Energy Experiment HETDEX survey \citep{hill08} will allow
$\approx750000$ Ly-$\alpha$ emitting galaxies to be mapped over
$1.9<z<3.5$ in 420deg$^2$ of sky in 150 clear nights. This survey is
claimed to measure the BAO scale to $\approx1$\% accuracy. This survey
will therefore produce similar errors to a QSO survey with an area of
4500deg$^2$ and a sky density of $\approx100$deg$^{-2}$, in a mostly
higher and hence  complementary redshift range.

\subsection{Markov Chain Monte Carlo cosmology fits} 

\begin{figure}
\epsfig{file=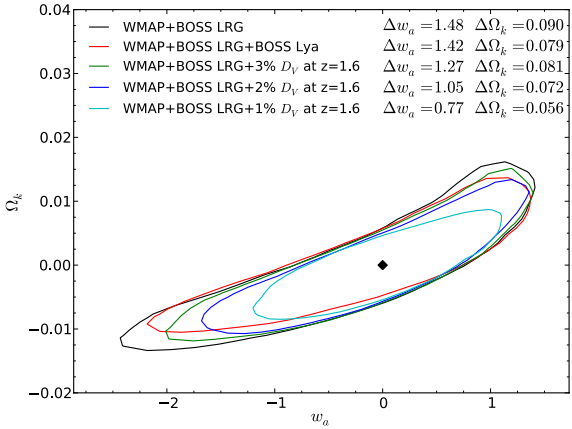,width=8.5cm}
\epsfig{file=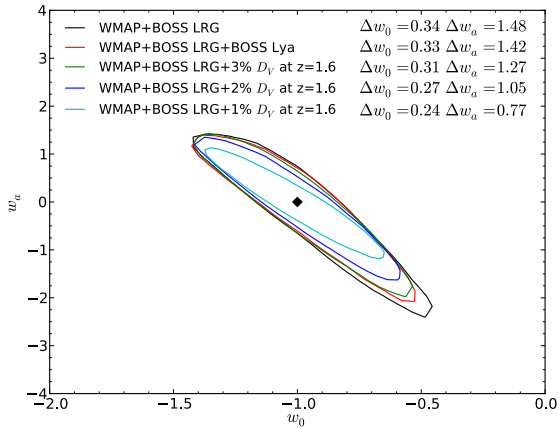,width=8.5cm}
\caption{\small Predicted survey constraints for $\Omega_k$, $w_a$,
$w_0$ from BAO MCMC fits. $1\sigma$ contours and errors are shown. {\bf
(a)} Our nominal QSO survey produces a 3\% BAO error (green contour)and
will provide the first detection of any non-standard (eg discontinuous)
evolution in the dark energy equation of state in the $1<z<2$ range. But
if the evolution remains relatively standard/continuous then a survey
with a 50\% bigger area and a 50\% higher sky density will  produce a
1-2\% QSO BAO error which will significantly improve over BOSS
constraints on $w_a$ and $\Omega_k$ (cyan, blue contours). {\bf (b)}
Similarly, even in the case of standard/continuous dark energy
evolution, a  survey with a 1-2\% BAO error will provide significantly
improved constraints over BOSS in the ($w_0$,$w_a$) plane (cyan, blue
contours).}
\label{fig:power2}
\end{figure}

Finally, we ran some MCMC cosmology fits for WMAP distance priors plus
current and future BAO surveys   for ($\Omega_m$, $\Omega_m h^2$, $w_0$,
$w_a$, $\Omega_k$) with the motivation of seeing if the data can
distinguish curvature from evolving dark energy.  For that reason we
first focus on the joint likelihood of ($w_a$, $\Omega_k$), which is
shown in Fig. \ref{fig:power2}a for various combinations of surveys.
We assume the CPL \citep{chev01,pol03} parameterisation for the evolving dark energy 
equation of state, i.e. $w(a)=w_0+w_a(1-a)$.
For the WMAP distance data we used the constraints on the shift parameter
$R$, the acoustic scale $l_a$ and the redshift of recombination $z_*$,
as given in \citet{Komatsu2008}.

Given just SDSS-LRG + WiggleZ (not shown), our nominal
3000deg$^2$/80deg$^2$ QSO survey at $z=1.6$ does give significant help.
However, Fig. \ref{fig:power2}a is based on WMAP $+$BOSS-LRG
($+$BOSS-Ly$\alpha$) and the accuracy of particularly the BOSS-LRG
measurement at $z\approx0.5$ provided by this combination means that the
QSOs (green contour) will only decrease the errors in $w_a, \Omega_k$ by
at most 10-20\% in this parameterization. Clearly in the case where
there is little evolution and $w_a\approx0$, even in their more
restricted redshift ranges, BOSS LRG and Ly-$\alpha$ surveys already
constrain $w_a$ as strongly   as the nominal 3000deg$^2$ +80deg$^{-2}$
QSO survey at $z\approx1.6$. The same result holds  in the $w_0,w_a$
plane in Fig. \ref{fig:power2}b. {\it On the other hand, it should be
noted that the nominal QSO survey would on its own produce the first
detection of, for example,  discontinuous dark energy evolution in the
so far unexplored $1<z<2.2$ redshift range.}

We  next  consider what survey parameters would lead to significant
improvements  in the cosmological forecasts, even in the case where the
dark energy evolves smoothly from $z\approx0.5$ to $z\approx1.6$. Bigger
QSO surveys shown in Fig. \ref{fig:power2}a,b assume 1\% and 2\% errors
in $D_V$ and the contours suggest that a 1.5\% error is likely to give
significantly smaller errors than the competing BOSS
LRG+Lyman-${\alpha}$ surveys in the $w_a, \Omega_k$ and the $w_0,w_a$
planes.

If so, then we would need to increase the QSO sky density by a factor of
$\approx1.6$ to $\approx140$deg$^{-2}$, reducing the QSO $D_V$ error by
a factor of $\approx1.5$ from 3\% to 2\%. This would require a 0.6mag
fainter mag limit taking us to $g<22.5$ rather than $g<21.85$. Then we
could drop the $z<1$ QSOs which takes us back to $\approx105$
deg$^{-2}$ (see Section 5.2), without much loss of BAO S/N as evidenced 
from the log-normal simulations and Fig. 4. This should be possible with 
VST ATLAS imaging data. If we wanted to get below a 2\% error then an additional
$\approx50$\% of the area ie $\approx4500$deg$^2$ would give a BAO error
of 1.6\%. This may be possible in a 1-2hr AAOmega exposure time and the
survey would then be completed in $\approx200$ clear nights. From Figs. 7, 
such a QSO survey would produce an error on the dark energy evolution
parameter, $w_a$, which is $\approx60$\% the size of that from BOSS
LRGs. The BAO detection significance would be $\approx6\sigma$ as opposed to
$3-4\sigma$ in the nominal 3000deg$^2$, 80deg$^{-2}$ QSO survey.

\subsection{Redshift-space distortions}
The nominal 3000deg$^2$  redshift survey of 250000 QSOs would also be able to
probe cosmology via redshift-space distortions. Now a redshift-space
distortion test of non-Einstein gravity is more difficult at high
redshift because $\Omega_m(z)$, at least in FLRW-based models, tends to 
unity at high z, making the $\gamma$ index of the gravitational growth
rate \citep{linder05}, $f(z)=\Omega_m(z)^\gamma$, more difficult to determine. However,
interesting cosmological constraints can still be obtained.  The infall
parameter governing redshift-space distortions, is defined as
$\beta=\Omega_m^{\gamma}/b$, and so depends on the bias, $b$, as well as
the gravitational growth rate, $\Omega_m^{\gamma}$. Previously for the
2QZ and 2SLAQ QSO surveys we have used redshift-space distortion and the
evolution of the QSO clustering amplitude to solve for $\Omega_m(z=0)$
and bias, $b(z=1.6)$ simultaneously \citep{jose05a,jose05b,jose08}. This
test also involves the Alcock-Pacynzski geometric test. The bias,
$b(z=1.6)$ can then be used to derive the amplitude of mass clustering,
ie $\sigma_8$, at $z=1.4$.

Recently, the combination $f \times \sigma_8^m$ ( $= \beta \times
\sigma_8^g$ if $b=\sigma_8^g/\sigma_8^m$) has become the prime target
for redshift-space distortion studies, since it can discriminate between
modified gravity models without needing to determine the bias
\citep{sp}. Redshift space distortions can thus provide a strong test of
Einstein's gravity independently of geometrical cosmological tests that
use standard candles and rods, such as BAO \citep{guzzo08}.
Redshift-space distortions can further be used to give an estimate of
the masses and hence mass-to-light ratios of galaxy group haloes in CDM
models (e.g. \citealt{gm09}).

\begin{figure}
\epsfig{file=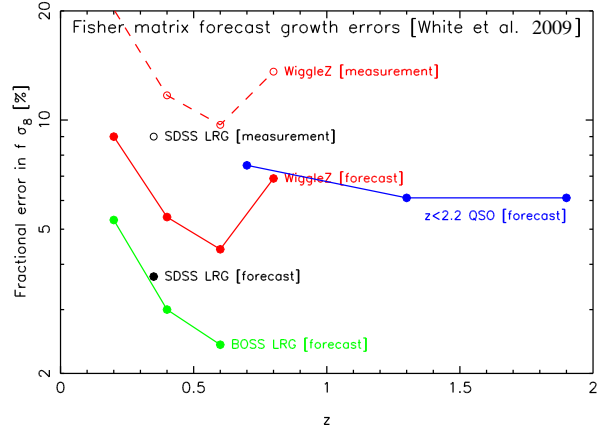,width=8.0cm}
\caption{\small Predicted sensitivity  for growth of structure in the
nominal QSO survey consistently compared  with other surveys.  
Growth of structure  forecasts are generated using the Fisher
matrix formula of \citet{wsp}.
}
\label{fig:rsd}
\end{figure}

In Fig. \ref{fig:rsd}, we show the error in the gravitational growth rate - mass
fluctuation parameter, $f\sigma_8$, for  the QSO survey as estimated
using the publicly available Fisher matrix  code of \citet{wsp}. The
result is averaged over the full range of clustering scales
$1<s<100$h$^{-1}$Mpc. We see that the overall  result is again
comparable to that for WiggleZ \citep{blake11} and the SDSS LRG survey but at a
significantly higher redshift. If we made a survey at the
$\approx140$deg$^{-2}$ QSO density and over an area of  4500deg$^2$, the error on
these redshift space distortion results would reduce by a further factor
of $\approx1.8\times$. So with these parameters the error on $f\sigma_8$ would
reduce from 6\% to 3.3\% in this redshift range making the survey even
more competitive in the $1<z<2.2$ range.

\section{QSO clustering as a probe of non-Gaussianity}
\subsection{Background}

Inflationary models with standard slow-roll inflation will produce  very
little non-Gaussianity, but models that deviate from the slow-roll
assumption can produce a significantly non-Gaussian seed field (see
\citealt{bartolo04} for a review). Non-Gaussianity is normally
parameterised through some amplitude of a quadratic term in the
primordial Bardeen potential, $f_{NL}$. This parameter gives the
coupling between a triangle of three k-modes, and can either be applied
to the local ('squeezed' isosceles triangles) or equilateral form of
non-Gaussianity. It can be measured using the large scale galaxy
Bispectrum. The best current limits on non-Gaussianity are obtained from
the cosmic microwave background radiation (CMB; \citealt{komatsu}) who
found $f_{NL}^{local}=32\pm21$. Galaxy surveys can actually be more
powerful as they can sample more modes in 3-D space than the CMB can on
the 2-D surface of the sphere. It also samples the structure of matter
perturbations at smaller scales, making it complementary to experiments
such as Planck. By probing scales in the range $k=0.01-0.2 h {\rm
Mpc}^{-1}$, it will be able to link constraints from the CMB with
constraints from clusters (\citealt{LoVerde2007}), constraining possible
scale-dependence of the non-Gaussianity.

In galaxy surveys, non-Gaussianity can produce a scale-dependent boost
of the halo power-spectrum at $k<0.03h$Mpc$^{-1}$ and this evolves as
$(1+z)$. Although this can potentially be confused with the full general 
relativistic correction of the galaxy power spectrum at $k \le 0.01h$Mpc$^{-1}$, 
the effect becomes important only beyond $z\approx3$ \citep{Yoo10}. 
Hence QSO surveys with their large volumes and redshift ranges 
make an ideal basis for this test \citep[e.g.][]{slosar08}. \citet{xia10a,xia10b} 
has recently produced upper limits on non-Gaussianity from the SDSS-DR6 
photo-z QSO catalogue \citep{richards09} and The NRAO VLA Sky Survey (NVSS; 
\citealt{condon98}). The NVSS auto-correlation function shows some
evidence for a positive tail extending to 5-6 degrees (\citealt{xia10a},
confirming previous results from \citealt{blake02}) which could be caused
by non-Gaussianity, implying $f_{NL}^{local}=62\pm27$. \citet{xia10b}
found lower but still consistent angular correlation functions from a
million QSOs in the SDSS DR6 dataset, implying $f_{NL}^{local}=58\pm24$. 
However, $f_{NL}$ measurements from high-$z$ photometric surveys can 
contain systematic bias due to gravitational lensing magnification 
\citep{Namikawa11}. QSO redshift surveys will provide more accurate and 
stringent constraints. 

\subsection{Non-Gaussianity constraints via the nominal QSO survey}

A 3000deg$^2$ QSO survey will  give highly competitive constraints on
primordial non-Gaussianity in the density field. \citet{sef07}
calculated how effective future galaxy surveys would be at
measuring the $f_{NL}$ parameters, simultaneously with the non-linear
bias. Their Fig. 6 shows predictions of the senstivity of surveys of
different volumes with a galaxy density of $5 \times 10^{-4}$
(h/Mpc)$^3$. Their forecasts demonstrate that our nominal QSO survey
has certain advantages over its low redshift counterparts. By making
measurements at higher redshift, it is less affected by non-linear
structure formation, and can measure the Bispectrum down to smaller
scales. Its main advantage will be its volume, which will be larger than
most other funded surveys planned for $z>1$. We estimate that our survey
will be able to constrain $f_{NL}$ (local) with an error of about $\pm15$,
and $f_{NL}$ (equilateral) of about $\pm150$, according to the 
\citet{sef07} analysis. This is better than any current survey :
these authors predict uncertainties almost $10\times$ larger for the SDSS-LRG
survey, for example. Our estimate is similar to the best current CMB
result, but this would be the first competitive test made using QSOs.

\subsection{Testing for non-Gaussianity via large-scale clustering evolution}

\begin{figure}
\epsfig{file=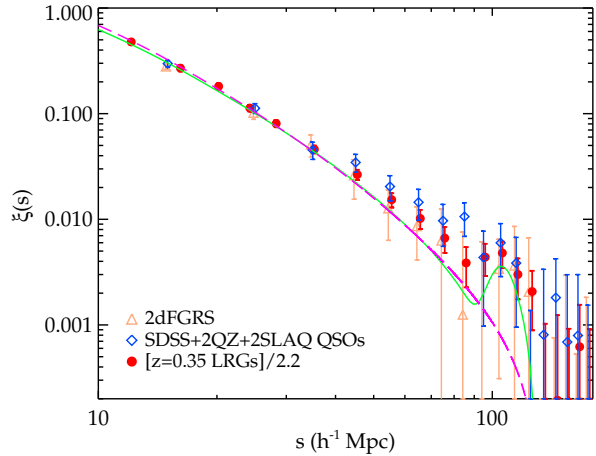,width=8.5cm}
\caption{\small The combined DR5$+$2QZ$+$2SLAQ QSO correlation function
from Figs. \ref{fig:xis_lin} now plotted on log axes and compared to the
2dFGRS $\xi(s)$ at $z=0.12$ of Hawkins et al and the scaled SDSS LRG
$\xi(s)$ of Eisenstein et al and a linear model fitted to LRG surveys in
the range $0.35<z<0.7$. The slope of the QSO $\xi(s)$ at $z\approx1.6$ is
marginally ($2\sigma$) flatter than the linear model fitted at $z<0.7$.}
\label{fig:logxi}
\end{figure}

Recently, angular correlation function studies of EROs in SA22 at
$z\approx1.5$ \citep{kim11} and LRGs  in SDSS Stripe 82 at $z\approx1$
\citep{nikos11} have shown evidence for a flatter slope in the range
$10<r<100$h$^{-1}$Mpc compared to the lower redshift,  SDSS
($z\approx0.35$), 2SLAQ ($z\approx0.55$) and AAOmega LRG
($z\approx0.68$) surveys. The evolution is small but statistically reasonably
significant ($\approx3\sigma$). Since evolution is not expected in the
linear regime in the standard cosmological model, one interpretation of
this evolution is that it might correspond to evidence for a
non-Gaussian feature in the galaxy correlation function becoming more
prominent at high redshifts, similar to that found above by \citet{xia10a,blake02}.

However, the angular correlation function is particularly susceptible to
artefacts in the data such as small artifical gradients. The amplitude
of the redshift space correlation function is intrinsically higher
because of the lack of projection effects. Therefore it is interesting
to use the QSO clustering correlation function, $\xi(s)$, to look for
evolution in the large-scale slope of the correlation function. In Fig.
\ref{fig:logxi} we therefore compare the combined QSO  correlation
function to the 2dFGRS $\xi(s)$ at $z\approx0.12$ \citep{hawkins03} and
the SDSS LRG $\xi(s)$ at $z\approx0.35$ \citep{Eisenstein05}. We also
show  a linear model that was fitted to correlation functions from 1.5
million LRGs in three photometric samples with average redshifts $z=
0.35, 0.55, 0.68$ \citep{sawangwitwtheta}. This model assumes a CDM
Universe with $\Omega_\Lambda=0.73$, $\Omega_m=0.27$, $f_{baryon}=
0.167$, $\sigma_8=0.8$, $h=0.7$ and $n_s=0.95$.  This model has also
been corrected for scale-dependent redshift-space distortion following
\citet{Eisenstein05}. It can be seen that this model also gives an
excellent fit to the 2dFGRS ($z=0.12$) and SDSS LRG ($z=0.35$) 3-D
correlation functions, $\xi(s)$, as shown. At the level of the current
errors in the observed correlation functions, it is not possible
statistically to distinguish the form of the observed low-$z$ galaxy and
high-$z$ QSO correlation functions. However, when $\chi^2$ fitted  in the 
$15<s<95$h$^{-1}$Mpc range, the slope of the
$z\approx1.6$ QSO correlation function appears  flatter  at the
$2\sigma$ level than the linear model fitted to the lower redshift
($z<0.7$) surveys.\footnote{Since the jack-knife errors are approximately 
Poisson in this range, we have ignored the covariance between points in our 
$\chi^2$ analysis.} This provides some limited support to the results
from the LRG angular correlation functions at $z=1$ and $z=1.5$ but
clearly more data are needed. Even the nominal QSO survey would provide
an $\approx3-4\times$ reduction in the errors in the 10-100h$^{-1}$Mpc
range and allow a much more significant search for non-Gaussian
evolution at large QSO separations. Furthermore, the possible detection
of non-Gaussian evolution in the current LRG and QSO surveys is a potent
reminder that dark energy evolution at high redshift may also be of an
unexpected form which should be measured rather than assumed.

\section{Other QSO z survey science}


More QSOs will also provide new data on the small-scale amplitudes of
QSO clustering as a function of luminosity and will improve $\xi(s)$
statistics at SDSS, 2QZ and 2SLAQ depths and this has importance for QSO
formation and evolution models. The amplitude of QSO clustering seems
remarkably independent of QSO luminosity \citep{shanks11}, is marginally 
significant (\citealt{porciani06} and \citealt{shen09}), and so checking for any small
luminosity dependence can put strict limits on models of QSO environment
and their host halo and BH mass.  A 3000deg$^2$ survey will contain
30000 QSOs at the SDSS depth, $\approx100000$ at the 2QZ flux limit and
$\approx120000$ at the 2SLAQ limit, so there will be significant improvements 
in the small-scale QSO clustering measurements in all 3 luminosity ranges.


A further application of a QSO survey is QSO lensing via magnification bias
(\citealt{myers03,myers05,scranton, menard}). Myers et al found
stronger  results using QSO spectroscopic $z$ than Menard et al did
using QSO photo-z. \citet{gm07} found that contamination of the photo-z
QSO sample by low redshift QSOs could reduce the anti-correlation signal
that is expected from lensing of faint QSOs. A large spectroscopic QSO 
survey would reduce the errors on the 2QZ results significantly and
test the validity of the photo-z results, as well as complementing  
galaxy shear weak-lensing tomography.


QSO surveys also allow investigations of the topology of the Universe.
For example, using the 2QZ survey, \citet{weatherley03} searched for
exact QSO spectral pairs at large separations to check for topologically
closed universes. Again a larger QSO survey would allow more stringent
constraints on such  models.

\section{Summary and Conclusions}

We first made a new correlation function analysis of the combined SDSS,
2QZ and 2SLAQ QSO surveys, comprising some 60000 QSOs. We focussed on
the large-scale, $s\approx100$h$^{-1}$Mpc, results to test the strength
of BAO signal that could be detected in the current dataset. We found
that the $\approx$22000 2QZ QSOs dominate the signal; although 2SLAQ has
a higher sky density it has only $\approx$6000 QSOs in total and
although SDSS has a larger number of QSOs its contribution is less
significant because of its low sky density. We observe a possible peak
at $\approx105$h$^{-1}$Mpc where the SDSS LRG $\xi(s)$ peak was found by
\citet{Eisenstein05} but here  it is only detected at a low significance
of $\approx1\sigma$ in the combined dataset and other peaks are 
seen at other separations at a similar significance.

We then proceeded to investigate the QSO survey parameters that would be
needed to make an $\approx4\sigma$ detection of the BAO peak. We conclude that
our nominal survey of 250000 QSOs in a sky area of 3000deg$^2$ will
allow us to make a $4\sigma$ detection of the BAO scale at $1<z<2.2$, an
as yet unexplored range for cosmology. This $\pm3$\% BAO scale
measurement will determine the high-redshift evolution of the dark
energy equation of state, $p=w(z)\rho$, and in particular show if there
is any large ($>15$\%) deviation from $w=-1$ in the $1<z<2.2$ redshift
range. But even if $w_a\approx0$, then a survey with 50\% higher QSO sky
density  and a 50\% bigger area will approximately halve the BAO error to 1.6\%. At
this point the QSO survey will also approximately halve the error on the dark
energy evolution  parameter, $w_a$, and significantly reduce the errors
on $w_0$ and $\Omega_k$ compared to the BOSS LRG and Lyman-$\alpha$ BAO
results at lower and higher redshifts.

A QSO survey can also set powerful new limits on the existence of
non-Gaussianity ($f_{NL}$) in the primordial density field. We have
found that the combined SDSS$+$2QZ$+$2SLAQ QSO survey shows possible
$\approx2\sigma$ evidence for evolution in the linear form of the
$z\approx1.6$ QSO correlation function, in the sense that it shows a
flatter slope than a linear model fitted to galaxy surveys at $z<0.7$.
Even the nominal QSO survey would improve the significance of detection
of this slope change by $3-4\times$. If confirmed, the result
could  indicate the detection of the evolution of a non-Gaussian
feature in the large-scale QSO correlation function. The tentative
evidence found here and in projected galaxy correlation functions for unexpected 
evolution of large-scale structure emphasises that dark energy 
evolution at high redshift may also be of an unexpected form.

The QSO survey will further support analyses of redshift space distortions
to measure $f\times\sigma_8$ and test Einstein gravity versus other 
modified gravity models. We shall also make the most rigorous
application of the \citet{ap79} test so far of the prediction of the
standard $\Lambda$CDM model for the amplitude of the mass clustering,
$\sigma_8$, at $z\approx1.5$. Other science it will do includes making
the most accurate determination of the luminosity dependence of QSO
clustering at small scales in order to probe QSO formation and evolution
via QSO environment. The survey could also use QSO lensing magnification
bias to measure the mass and the bias of foreground groups and clusters
and to do lensing tomography.

\section*{Acknowledgments}
We acknowledge Chris Blake for allowing us to access the results from
log-normal simulations of \citet{blakekgb03, blake06} and publicly available
Fisher matrix codes. We thank the referee for useful comments which helped 
improve the paper.

US acknowledges financial support from the Institute for the
Promotion of Teaching Science and Technology (IPST) of
The Royal Thai Government. 

We  thank all the present and former staff of the 
Anglo-Australian Observatory for their work in building and operating
the 2dF facility. 

Funding for the SDSS and SDSS-II has been provided by the Alfred
P. Sloan Foundation, the Participating Institutions, the National
Science Foundation, the U.S. Department of Energy, the National
Aeronautics and Space Administration, the Japanese Monbukagakusho, the
Max Planck Society, and the Higher Education Funding Council for
England. The SDSS Web Site is {\tt http://www.sdss.org/}.

\setlength{\bibhang}{2.0em}
\setlength\labelwidth{0.0em}


\label{lastpage}

\end{document}